\documentclass[preprint,numbers,9pt]{sigplanconf}

\usepackage{booktabs}
\usepackage{amsmath,amssymb}
\usepackage{url}
\usepackage{listings}
\usepackage{algorithm}
\usepackage{algpseudocode}
\usepackage{xspace}
\usepackage{times}
\usepackage{subcaption}
\usepackage{framed}
\usepackage{hyperref}
\usepackage{balance}
\usepackage[T1]{fontenc}
\usepackage[utf8]{inputenc}
\usepackage{microtype}
\usepackage{todonotes}
\usepackage{adjustbox}
\usepackage{graphicx}

\algrenewcommand\algorithmicindent{1.0em}%
\algnewcommand\algorithmicforeach{\textbf{for each}}
\algdef{S}[FOR]{ForEach}[1]{\algorithmicforeach\ #1\ \algorithmicdo}

\newcommand{\eg}{\emph{e.g.}\xspace}
\newcommand{\ie}{\emph{i.e.}\xspace}

\newcommand{\taskprof}{\textsc{TaskProf}\xspace}

\begin{document}
\title{A Fast Causal Profiler for Task Parallel Programs}

\authorinfo{Adarsh Yoga}{Rutgers University}{adarsh.yoga@cs.rutgers.edu}
\authorinfo{Santosh Nagarakatte}{Rutgers University}{santosh.nagarakatte@cs.rutgers.edu}

\maketitle

\begin{abstract}
This paper proposes \taskprof, a profiler that identifies parallelism
bottlenecks in task parallel programs. It leverages the structure of a
task parallel execution to perform fine-grained attribution of work to
various parts of the program. \taskprof's use of hardware performance
counters to perform fine-grained measurements minimizes perturbation.
\taskprof's profile execution runs in parallel using multi-cores.
\taskprof's causal profile enables users to estimate improvements in
parallelism when a region of code is optimized even when concrete
optimizations are not yet known.  We have used \taskprof to isolate
parallelism bottlenecks in twenty three applications that use the
Intel Threading Building Blocks library.  We have designed
parallelization techniques in five applications to increase
parallelism by an order of magnitude using \taskprof. Our user study
indicates that developers are able to isolate performance bottlenecks
with ease using \taskprof.
\end{abstract}

\section{Introduction}
\begin{figure*}
  \includegraphics[width=\textwidth]{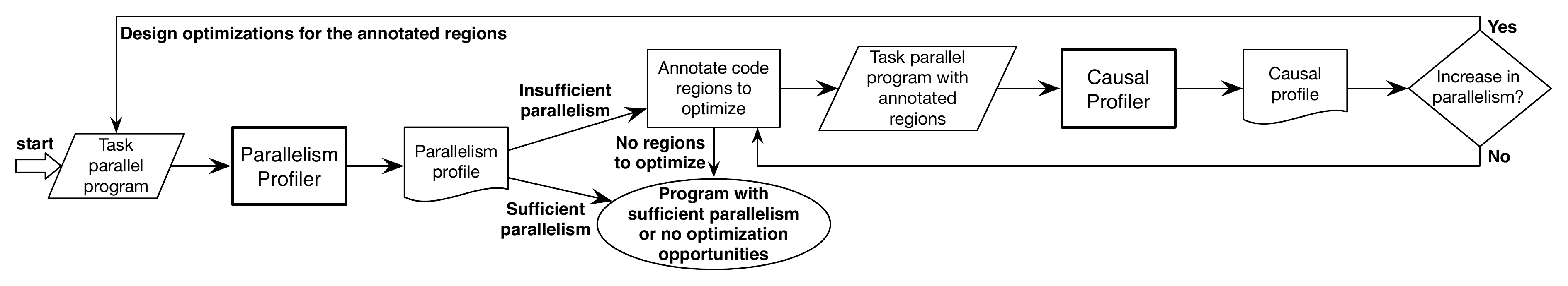}
  \caption{\small Identifying and diagnosing parallelism bottlenecks
    in task parallel programs using \taskprof's parallelism and causal
    profiles.}
  \label{fig:workflow}
\end{figure*}

Task parallelism is an effective approach to write performance
portable code~\cite{Grossmann:2012}. In this model, the programmer
specifies fine-grained tasks and the runtime maps these tasks to
processors while automatically balancing the workload using work
stealing algorithms.  Many task parallelism frameworks have become
mainstream (\eg, Intel Threading Building Blocks
(TBB)~\cite{Reinders:2007}, Cilk~\cite{Frigo:1998}, Microsoft Task
Parallel Library~\cite{Leijen:2009}, Habanero Java~\cite{Cave:2011},
X10~\cite{Charles:2005}, and Java Fork/Join tasks~\cite{Lea:2000}).

A common metric used to quantify the performance of a task parallel
program is asymptotic parallelism, which measures the potential
speedup when the program is executed on a large number of processors.
It is constrained by the longest chain of tasks that must be executed
sequentially~(also known as the \texttt{span} or the critical
work). Hence, asymptotic parallelism is the ratio of the total work
and the critical work performed by the program for a given input. A
scalable program must have large asymptotic parallelism.
A task parallel program can have low asymptotic parallelism due to
multiple factors: coarse-grained tasks, limited work performed by the
program, and secondary effects of execution such as contention, low
locality, and false sharing.

Numerous techniques have been proposed to address various bottlenecks
in both multithreaded programs~\cite{Tallent:2010:ALC, Chen:2012,
  DuBois:2013:CSI, Yu:2016, David:2014, Liu:2013:DPP, Liu:2015,
  Liu:2013:PDL} and task parallel
programs~\cite{Schardl:2015,He:2010}. These technique range from
identifying critical paths~\cite{Hollingsworth:1994, Miller:1990,
  Oyama:2000}, parallelism~\cite{Schardl:2015,He:2010},
synchronization bottlenecks~\cite{Tallent:2010:ALC, Chen:2012,
  DuBois:2013:CSI, Yu:2016, David:2014}, and other performance
pathologies~\cite{Liu:2013:DPP, Liu:2015, Liu:2013:PDL}.
Tools for multithreaded programs identify bottlenecks in a specific
execution on a specific machine, which does not necessarily provide
information about scalability of the program. In contrast, tools that
measure asymptotic parallelism in task parallel programs run the
program serially~\cite{Schardl:2015,He:2010}, which is feasible only
when the task parallel model provides serial semantics (\eg,
Cilk)~\cite{Frigo:1998}.
Although they identify parallelism bottlenecks, they do not provide
information on regions of code that matter in improving asymptotic
parallelism. 

This paper proposes \taskprof, a fast and causal profiler that
measures asymptotic parallelism in task parallel programs for a given
input. \taskprof's causal profile allows users to estimate
improvements in parallelism when regions of code are optimized even
before concrete optimizations for them are known.  \taskprof has three
main goals: (1) to minimize perturbation~(also known as
interference~\cite{Graham:1983}) while accurately computing asymptotic
parallelism and critical work for each spawn site (source code
location where a task is created), (2) to run the profiler in
parallel, and (3) to provide feedback on regions of code that matter
in increasing parallelism.

\taskprof computes an accurate parallelism profile by performing a
fine-grained attribution of work to various parts of the program using
the structure of a task parallel execution.
The execution of a task parallel program can be represented as a
tree~(specifically Dynamic Program Structure
Tree~(DPST)~\cite{Raman:2012}), which captures the series-parallel
relationships between tasks and can be constructed in parallel.  Given
a task parallel program, \taskprof constructs the DPST in parallel
during program execution and attributes work to leaves of the DPST. To
minimize perturbation, \taskprof uses hardware performance counters to
measure work performed in regions without any task management
constructs, which correspond to leaves in the DPST. \taskprof writes
the DPST and the work performed by the leaf nodes of the DPST to a
profile data file.  The profile execution runs in parallel leveraging
multi-cores and the measurement of computation using performance
counters is thread-safe.

\taskprof's post-execution analysis tool uses the data file from the
profile run, reconstructs the DPST, and computes asymptotic
parallelism and critical work at each spawn site in the program using
the properties of the DPST (see Section~\ref{sec.offline.analysis}).
\taskprof maps dynamic execution information to static spawn sites by
maintaining information about spawn sites in the DPST.  \taskprof's
profile for the sample program in Figure~\ref{fig:example} is shown in
Figure~\ref{fig:dpst}(b).

The spawn sites that perform a large fraction of the critical work in
the profile are the parallelism bottlenecks in the program.  However,
optimizing regions that perform critical work may not increase
asymptotic parallelism when the program has multiple regions that
perform similar amount of critical work.  Designing a parallelization
strategy that reduces critical work requires significant
effort. Hence, the programmer would like to know if optimizing a
region of code increases asymptotic parallelism even before the
specific optimization is designed.

%
%

\taskprof provides a causal profile that estimates the improvement in
asymptotic parallelism when a specific region of code in the program
is optimized even before concrete optimizations for them are
known. \taskprof's causal profile is inspired by
\textsc{Coz}~\cite{Curtsinger:2015} that quantifies the speedup when a
selected program fragment is optimized in multithreaded programs by
slowing down all code executing concurrently with the
fragment. However, \textsc{Coz} cannot be used with task parallel
programs as it is not possible to slow down all active tasks.

In contrast, \taskprof is able to generate a causal profile because it
builds an accurate performance model of a task parallel execution by
performing a fine-grained attribution of work to the nodes of the
DPST. To quantify the impact of optimizing a region of code, the
programmer annotates the beginning and the end of the region in the
program and the anticipated speedup for the region. \taskprof
generates a causal profile that shows the increase in parallelism with
varying amounts of anticipated speedup for the annotated regions (see
Figure~\ref{fig:dpst}(c)).
To generate a causal profile, \taskprof re-executes the program,
generates profile data, and identifies nodes in the DPST that
correspond to the annotated regions. Subsequently, \taskprof
recomputes the asymptotic parallelism in the program 
by reducing the critical work of the annotated region of code by the
anticipated improvement.  \taskprof's causal profiling enables the
programmer to identify improvements in asymptotic parallelism even
before the developer actually designs the
optimization. Figure~\ref{fig:workflow} illustrates \taskprof's usage
to generate a parallelism profile and a causal profile.

\taskprof prototype is open source and available
online~\cite{taskprof-git}.  We have identified parallelism
bottlenecks in twenty three Intel TBB applications using the
prototype.
Using \taskprof's causal profile, we also designed concrete
parallelization techniques for five applications to address the
parallelism bottlenecks.
Our concrete optimizations increased parallelism in these five
applications by an order of magnitude.  We conducted a user study
involving thirteen undergraduate and graduate students to evaluate the
usability of \taskprof. Our results show that the participants quickly
diagnosed parallelism bottlenecks using \taskprof.


\section{Background}
This section provides a quick primer on the tree-based representation
of a task parallel execution, which is used by \taskprof to compute
parallelism and causal profiles.

\vspace{4pt} \textbf{Task parallelism.} Task parallelism is a
structured parallel programming model that simplifies the job of
writing performance portable code. In this model, parallel programs
are expressed using a small set of expressive yet structured patterns.
In contrast to threads, task creation is inexpensive and a task is
typically bound to the same thread till completion~\cite{McCool:2012}.
The runtime uses work stealing to map dynamic tasks to runtime threads
and balances the workload between threads~\cite{Frigo:1998}. Task
programming models provide specific constructs to create tasks~(\eg,
\texttt{spawn} keyword in Cilk and \texttt{spawn} function in Intel
TBB) and to wait for other tasks to complete~(\eg, \texttt{sync}
keyword in Cilk and \texttt{wait\_for\_all()} function in Intel TBB).
A sample task parallel program is shown in
Figure~\ref{fig:example}. These models also provide patterns for
recursive decomposition of a program~(\eg, \texttt{parallel\_for} and
\texttt{parallel\_reduce}) that are built using the basic constructs.
Task parallelism is expressive and widely applicable for writing
structured parallel programs.

\begin{figure}
\lstset{
  language=C++,
  basicstyle=\ttfamily\footnotesize,
  numbers=left,
  tabsize=2,
  breaklines=true,
  captionpos=b,
  xleftmargin=2.5em,
  numberblanklines=false,
  morekeywords={__CAUSAL_BEGIN__,__CAUSAL_END__}
}
\begin{lstlisting}
void compute_tree_sum(node* n, int* sum) {
  if(n->num_nodes <= BASE) {
    //Compute sum serially
    __CAUSAL_BEGIN__
    *sum = serial_tree_sum(n);
    __CAUSAL_END__
  } else {
    int left_sum, right_sum;
    if(n->left) {
      spawn compute_tree_sum(n->left, &left_sum);
    }
    if(n->right) {
      spawn compute_tree_sum(n->right, &right_sum);
    }
    sync;
    *sum = left_sum + right_sum;
  }
}
int main() {
  __CAUSAL_BEGIN__
  node* root = create_tree();
  __CAUSAL_END__
  int sum;
  spawn compute_tree_sum(root, &sum);
  sync;
  //print sum;
  return 0;
}
\end{lstlisting}
\caption{A program that computes the sum of the nodes in a binary
  tree. It creates tasks and waits for tasks to complete using
  \texttt{spawn} and \texttt{sync} keywords, respectively. Each node
  in the tree holds an integer value, number of nodes in the sub-tree
  rooted at the node, and pointers to the left and right sub-tree.
  The \texttt{create\_tree} function builds the tree.  The
  \texttt{serial\_tree\_sum} takes a node \texttt{n} as argument and
  computes the sum in the sub-tree under \texttt{n}. BASE is a
  constant that determines the amount of serial work.  The user has
  used annotations (\texttt{\_\_CAUSAL\_BEGIN\_\_} and
  \texttt{\_\_CAUSAL\_END\_\_}) to specify regions for causal
  profiling, which are not used in the regular profiling phase.}
\label{fig:example}
\end{figure}

\begin{figure}
  \includegraphics[width=\linewidth]{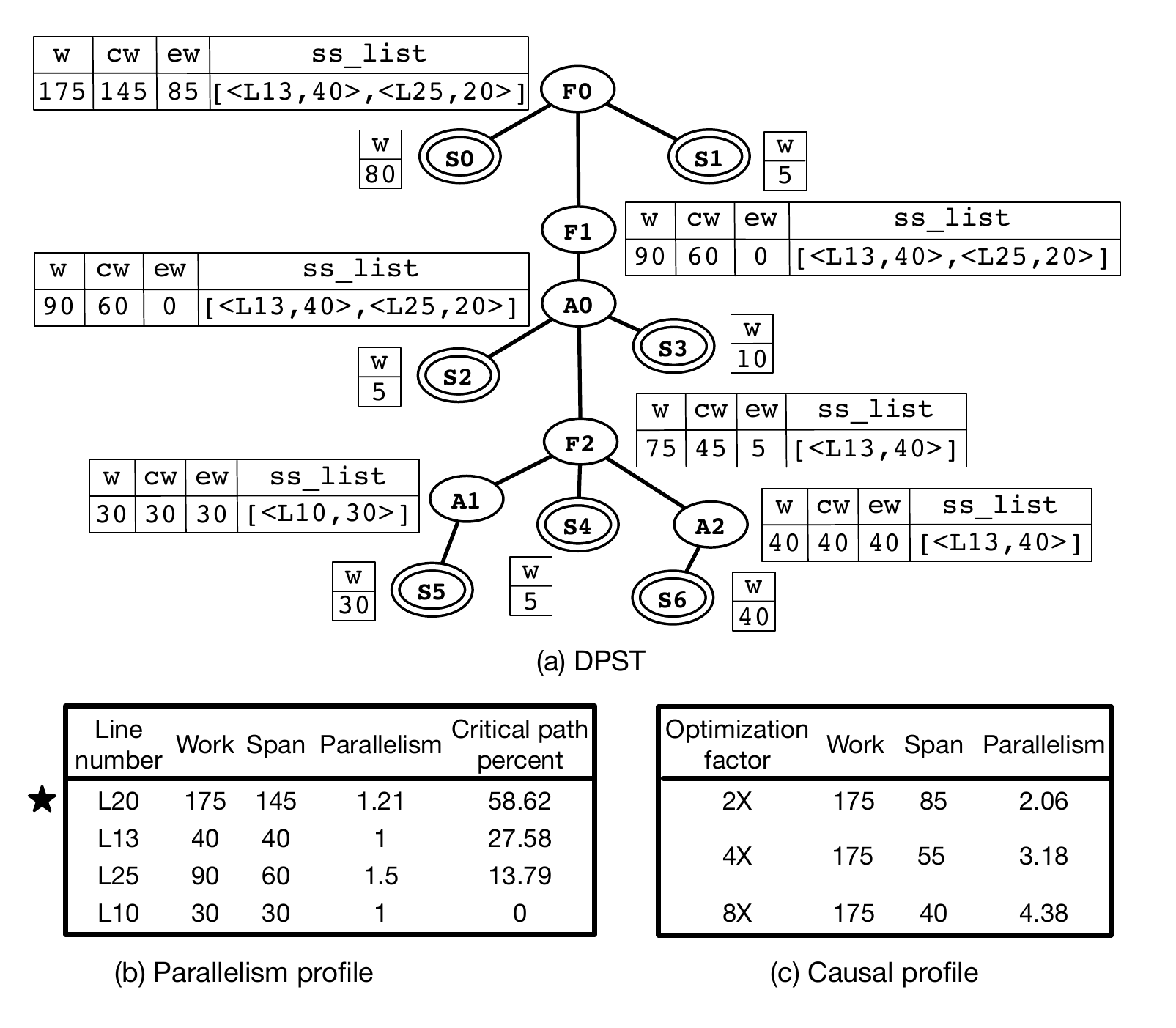}
  \caption{(a) The DPST for an execution of the program in
    Figure~\ref{fig:example}. $F0$, $F1$, and $F2$ are finish
    nodes. $A0$, $A1$, and $A2$ are async nodes. Step nodes are leaves
    in the DPST.  \taskprof maintains four quantities with each
    intermediate node in the DPST: work~(\texttt{w}), critical
    work~(\texttt{cw}), exclusive work~(\texttt{ew}), and the list of
    spawn sites performing critical work~(\texttt{ss\_list}). Each
    entry in the spawn site list maintains the line number and the
    exclusive work done by the spawn site~(\eg, $<L20, 40>$). Step
    nodes have work data from the profile execution. \taskprof updates
    these quantities for the intermediate nodes by performing a
    bottom-up traversal of the DPST.  (b) The profile generated by
    \taskprof reports the work, critical work, parallelism, and
    percentage of critical work with each spawn site. Line with a
    ``$\star$'' in the profile corresponds to the main function and
    reports the parallelism for the entire program.  (c) The causal
    profile reports the parallelism for the whole program when the
    annotated regions in Figure~\ref{fig:example} are optimized by
    2$\times$, 4$\times$, and 8$\times$.}
  \label{fig:dpst}
\end{figure}

\vspace{4pt}

\vspace{4pt}
\textbf{Dynamic program structure tree.} The execution of a task
parallel program can be represented as a dynamic program structure
tree (DPST), which precisely captures the series-parallel
relationships between tasks~\cite{Raman:2012}. Further, the DPST can
be constructed in parallel.  Since our goal in this paper is to
profile the program in parallel, we use the DPST representation of a
task parallel execution.

The DPST is a n-ary tree representation of a task parallel execution.
There are three kinds of nodes in a DPST: (1) \texttt{step}, (2)
\texttt{async}, and (3) \texttt{finish} nodes.  The step node
represents the sequence of dynamic instructions without any task spawn
or sync statements.  All computations occur in step nodes. The
async node in the DPST represents the creation of a child
task by a parent task. The descendants of the newly created task can
execute in parallel with the remainder of the parent task.  A finish
node is created in a DPST when a task spawns a child task and waits
for the child (and its descendants) to complete. A finish node is the
parent of all async, finish and step nodes directly executed by its
children or their descendants.

The DPST, by construction, ensures that two parallel tasks operate on
two disjoint sub-trees.  DPST's construction also ensures that all
internal nodes are either async or finish nodes.  The siblings of a
particular node in a DPST are ordered left-to-right to reflect the
left-to-right sequencing of computation of their parent task.
A path from a node to the root and the
left-to-right ordering of siblings in a DPST do not change even when
nodes are added to the DPST during execution.
The DPST was originally used for data race detection because it allows
a race detector to check if two accesses can occur in
parallel~\cite{Yoga:2016:PDR, Raman:2012}.  In a DPST, two step nodes
$S1$ and $S2$ (assuming $S1$ is to the left of $S2$) can execute in
parallel if the least common ancestor of $S1$ and $S2$ in the DPST has
an immediate child that is an async node and is also an ancestor of
$S1$. In Section~\ref{sec.offline.analysis}, we will highlight the
properties of the DPST that we use to profile programs.

\vspace{4pt}
\textbf{Illustration of the DPST.}  Figure~\ref{fig:dpst}(a) shows the
DPST for an execution of the program in Figure~\ref{fig:example}. The
program in Figure~\ref{fig:example} will execute the spawn call at
line 10 and line 13 once when BASE=n/2, where $n$ is the number of
nodes in the tree.

We construct the DPST during program execution as follows.  When the
main function starts, we add a finish node $F0$ as the root of the
DPST to represent the fact that main completes after all the tasks
spawned by it have completed. We add a step node $S0$ as the child of
the root finish node to capture the initial computations being
performed in the main function. On a spawn call at line 24 in
Figure~\ref{fig:example}, we create a finish node $F1$ because it is
the first spawn performed by the task. We also add an async node $A0$
as the child of $F1$ to represent the spawning of a task. Any
computation by the newly created task will be added as nodes in the
sub-tree under the async node $A0$. The operations performed in the
continuation of the main task will be added to the right of the async
node $A0$ under the finish node $F1$.  Hence, the continuation of the
main task and newly created task operate on distinct subtrees of the
DPST and can update the DPST in parallel.

\section{Parallelism Profiler}
\label{sec.profile}

\taskprof computes the total work, part of the total work done
serially (critical work or span), and the asymptotic parallelism at
each spawn site in a task parallel program. The key contribution of
\taskprof is in fine-grained attribution of work while ensuring that
the profile execution is fast, perturbation-free, and
accurate. \taskprof accomplishes the goal of fast profile execution by
using multi-cores. \taskprof's profile execution itself runs in
parallel and leverages the DPST representation to attribute work to
various parts of the program.  \taskprof ensures that the profile
execution is perturbation-free by using hardware performance counters
to obtain information about the computation performed by the step
nodes in the DPST. \taskprof also maintains a very small fraction of
the DPST in memory during profile execution to further minimize
perturbation. \taskprof ensures that the parallelism profile is
accurate by capturing spawn sites through compiler instrumentation and
by precisely measuring work performed in each step node.

\taskprof computes the parallelism profile in three steps. First,
\taskprof provides a modified library for task parallelism that
captures information about spawn sites. \taskprof's compiler
instrumentation modifies the calls to the task parallel library in the
program to provide information about spawn sites.
Second, \taskprof's profile execution runs in parallel on the
multi-core processors, constructs the DPST representation of the
execution, and collects fine-grained information about the execution
using hardware performance counters. \taskprof writes the profile
information to a data file similar to the \texttt{grof} profiler for
sequential programs~\cite{Graham:1982}.
Third, \taskprof's offline analysis tool analyzes the profile data and
aggregates the data for each static spawn site. Finally, it computes
asymptotic parallelism and critical work for each spawn site.

\vspace{4pt} \textbf{Static instrumentation and modified libraries.}
\taskprof provides a modified task parallelism library that constructs
the DPST and reads hardware performance counters at the begining and
end of each step node.  \taskprof uses static instrumentation to
instrument the program with calls to the modified task parallel
runtime library.
In the subsequent offline analysis phase, \taskprof needs to map the
dynamic execution information about asymptotic parallelism and
critical work to static spawn sites in the program. Hence, \taskprof
instruments the spawn sites to capture the line number and the file
name of the spawn site. \taskprof's static instrumentation is
currently structured as a rewriter over the abstract syntax tree of
the program using the Clang compiler front-end. Our instrumented
libraries and compiler instrumentation enable the programmer to use
\taskprof without making any changes to the source code.

\subsection{Parallel Profile Execution}
The goal of the profile execution is to collect fine-grained
information about the program to enable a subsequent offline
computation of asymptotic parallelism. Typically, programs are
profiled with representative production inputs that have long
execution times. Hence, a fast profile execution is desirable.
Our goal is to ensure that the execution time of the program with and
without profiling is similar. Hence, \taskprof profiles in parallel
leveraging multi-core processors.
To ensure a parallel profile execution, it needs to construct the
execution graph in parallel and collect information about the program
in a thread-safe manner.

\vspace{4pt}
\textbf{The DPST representation for parallel profile execution.} We
use the DPST representation to measure work performed by various parts
of the program because the DPST can be constructed in parallel.
\taskprof constructs the DPST as the program executes the injected
static instrumentation and measures the work performed in each step
node. The DPST, once constructed, allows \taskprof to determine the
dependencies between tasks. This fine-grained attribution of work to
the step nodes in the DPST enables \taskprof to compute the
parallelism in the program eventually using an offline analysis.

The DPST of the complete task parallel execution has a large number of
nodes. Storing the entire DPST in memory during program execution can
cause memory overheads and perturb the execution.  To address this
issue, \taskprof does not maintain the entire DPST in memory. In a
library based task parallel programming model, a task is always
attached to the same thread.
We leverage this property to minimize the footprint of the DPST in
memory.
\taskprof maintains a small fraction of the nodes that correspond to
the tasks currently executing on each thread in memory.

Once a step node of a task completes execution, the work performed in
the step node along with the information about its parent node is
written to the profile data file and the DPST node can be deallocated.
As async nodes do not perform any work, \taskprof writes the
information about its parent in the DPST and the spawn site associated
with the async node to the profile data file. In contrast to step and
async nodes, only parent node information is written to the profile
data file for a finish node.  

\vspace{4pt}
\textbf{Measuring work with hardware performance counters.} To measure
the work performed in each step node without performance overhead,
\taskprof uses hardware performance counters. Performance counters are
model specific registers available that count various events performed
by the hardware using precise event-based sampling
mechanisms~\cite{Intel:2016}. These performance counters can be
programmatically accessed.  \taskprof can use both the number of
dynamic instructions and the number of execution cycles to measure the
work done in a step node. Measuring execution cycles allows \taskprof
to account for latencies due to secondary effects such as locality,
sharing, and long latency instructions.
Further, the operations on these counters are thread-safe. \taskprof
reads the value of the counter at the beginning and the end of the 
step node using static instrumentation injected into the program. It
calculates the work performed in the step node by computing difference
between the two counter values.  This fine-grained measurement of work
performed in each step node using hardware performance counters along
with the construction of the DPST while executing in parallel allows
\taskprof to compute a precise, yet fast profile of the program.

The profile data file generated at the end of parallel profile
execution contains the work done in each step node. It also contains
the information about the parent for each node in the DPST and the
spawn site information for each async node. The left-to-right
sequencing of nodes is implicitly captured by the order of the nodes
in the profile data file.

\subsection{Offline Analysis of the Profile Data}
\label{sec.offline.analysis}
\taskprof's offline analysis reconstructs the DPST using the data from
the profile execution and computes the work and critical work (span)
for each spawn site in the program. The construction of the DPST from
the profile data is fairly straightforward as it contains information
about nodes, their parent nodes, and the left-to-right ordering of the
nodes. In this section, we describe the computation of work and span
for each intermediate node in the DPST given the work performed in the
step nodes. We also describe the process of mapping this dynamic
information to static spawn sites.

\begin{figure}
\begin{algorithmic}[1]
\Function{ComputeWorkSpan}{$T$}
  \ForEach{intermediate node $N$ in bottom-up traversal of $T$}
    \State $C_N \gets \Call{Children}{N}$
    \State $N.work \gets \displaystyle\sum_{C \in C_N} C.work$
    \State $S_N \gets \Call{StepChildren}{N}$
    \State $F_N \gets \Call{FinishChildren}{N}$
    \State $N.c\_work \gets \displaystyle\sum_{S \in S_N} S.work + \displaystyle\sum_{F \in F_N} F.c\_work$
    \State $N.e\_work \gets \displaystyle\sum_{S \in S_N} S.work + \displaystyle\sum_{F \in F_N} F.e\_work$
    \State $N.ss\_list \gets \displaystyle\bigcup_{F \in F_N} F.ss\_list$

    \ForEach{$A \in \Call{AsyncChildren}{N}$}
      \State $LS_A \gets \Call{LeftStepSiblings}{A}$
      \State $LF_A \gets \Call{LeftFinishSiblings}{A}$
      \State $llw_A \gets \displaystyle\sum_{LS \in LS_A} LS.work + \displaystyle\sum_{LF \in LF_A} LF.c\_work$
      \If{$llw_A + A.c\_work > N.c\_work$}
        \State $N.c\_work \gets llw_A + A.c\_work$
        \State $N.e\_work \gets \displaystyle\sum_{S \in LS_A} S.work+\displaystyle\sum_{F \in LF_A} F.e\_work$
        \State $N.ss\_list \gets (\displaystyle\bigcup_{LF \in LF_A} LF.ss\_list) \cup A.ss\_list$
      \EndIf
    \EndFor

    \If{$N$ is a \texttt{async} node}
      \State $N.ss\_list \gets N.ss\_list \cup \langle N.s\_site, N.e\_work \rangle$
    \EndIf
  \EndFor
\EndFunction
\end{algorithmic}
\caption{Algorithm to compute the total work (\texttt{work}),
  critical work (\texttt{c\_work}), exclusive work (\texttt{e\_work}),
  and the spawn sites that perform the critical
  work~(\texttt{ss\_list}) for each intermediate node in the DPST.
  The function \protect\Call{Children}{} returns the set of children
  of the input node. Similarly, functions
  \protect\Call{StepChildren}{}, \protect\Call{FinishChildren}{} and
  \protect\Call{AsyncChildren}{} return the set of \texttt{step},
  \texttt{finish} and \texttt{async} child nodes of the input node,
  respectively. The function \protect\Call{LeftStepSiblings}{} returns
  the set of \texttt{step} sibling nodes that occur to the left of the
  input node in the DPST. Similarly, the
  \protect\Call{LeftFinishSiblings}{} returns the set of
  \texttt{finish} sibling nodes to the left of the input node in the
  DPST.}
\label{alg.workspan}
\end{figure}

\vspace{4pt}
\textbf{Computing work and critical work for each intermediate node.}  
In the DPST representation, all computation is performed in the step
nodes. The step nodes have fine-grained work information from the
profile execution. \taskprof needs to compute the total work and the
fraction of that work done serially (critical work) for each
intermediate node in the DPST. To provide meaningful feedback to the
programmer, \taskprof also computes the list of spawn sites that
perform critical work and the portion of the critical work performed
exclusively by each spawn site.

\taskprof computes the total work and the critical work at each
intermediate node by performing a bottom-up traversal of the DPST. The
total work performed in the sub-tree at each intermediate node is sum
of the work performed by all the step nodes in the sub-tree.
In contrast, critical work measures the amount of work that is
performed serially. Computing critical work and the set of tasks
performing the critical work requires us to leverage the properties of
the DPST. Specifically, we leverage the following properties of the
DPST to compute the critical work.
\begin{itemize}

\item The siblings of a node in a DPST are ordered left-to-right
  reflecting the left-to-right sequencing in the parent task.
\item Given an intermediate node, all the direct step children of the node
  execute serially.
\item All the left step or finish siblings of an async node execute
  serially with the descendants of the async node.
\item All the right siblings (and their descendants) of an async node
  execute in parallel with the descendants of the async node.
\end{itemize}

Using the above properties of the DPST, the critical work at an
intermediate node will be equal to either (1) the serial work done by
all the direct step children and the critical work performed by the
finish children or (2) the critical work performed by descendants of
an async child and the serial work performed by the left step and
finish siblings of the specific async child in consideration.  Since
any intermediate node in the DPST can have multiple async children,
\taskprof needs to check if any of the async nodes can contribute to
the critical work. For example, consider the intermediate node $F2$
in Figure~\ref{fig:dpst}(a) that has two async nodes $A1$ and
$A2$. The critical work will be the maximum of (1) the work
done by the direct step child $S4$ or (2) the critical work
 by the async child $A1$ (it does not have any left
siblings), or (3) the sum of the critical work by the async child $A2$ and
the work done by the step node $S4$, which is the left step sibling of $A2$.

%
Each async node in the DPST corresponds to a spawn site in the program
because async nodes are created when a new task is spawned. Hence,
\taskprof computes the list of spawn sites performing critical work by
computing the list of async nodes that contribute to the critical work
in the sub-tree of the intermediate node.





\vspace{4pt} \textbf{Algorithm to compute work and critical work.}
Figure~\ref{alg.workspan} provides the algorithm used by \taskprof to
compute the total work, the critical work, and the set of spawn sites
contributing to the critical work.
The algorithm maintains four quantities with each intermediate node in
the DPST: (1) total work performed in the sub-tree under the node
(\texttt{work}), (2) the critical work performed in the sub-tree
~(\texttt{c\_work}), (3) the list of spawn sites that perform the
critical work~(\texttt{ss\_list}), and (4) the part of the critical
work that is performed exclusively by the direct children of the
node (\texttt{e\_work}).
The exclusive work of a node is equal to sum total of the work
performed by the direct step children and the exclusive work performed
by the finish children. We consider the exclusive work performed by a
finish node because it is not yet associated with any spawn site. The
exclusive work of the current node will eventually be associated with
a spawn site.  The algorithm does not consider the exclusive work of
the async children because it is already associated with a spawn site.

The algorithm traverses each node in the DPST in a bottom up fashion.
All step nodes have work information from the profile data. For any
intermediate node, the work performed under the sub-tree is the sum of
the work performed by all its children (lines 3-4 in
Figure~\ref{alg.workspan}).
For a given intermediate node, \taskprof initially computes the serial
work performed in all the step and finish children as the critical
work~(lines 5-7 in Figure~\ref{alg.workspan}). For each async child of
the current node, it checks if the serial work done by the async node
and its left siblings is greater than the critical work computed until
that point~(lines 10-15 in Figure~\ref{alg.workspan}).


To compute the set of spawn sites performing critical work, each
intermediate node also maintains a list of spawn sites and the
exclusive work performed by them. The algorithm initially sets the
spawn site list for a node to be the union of spawn site lists of its
finish children~(lines 8-9 in Figure~\ref{alg.workspan}).
Whenever an async child contributes to the critical work, the spawn site
list of the current node is the union of the spawn site list of the
async child and the spawn site lists of the finish children that are
to the left of the async child~(line 17 in Figure~\ref{alg.workspan}).
When an async child contributes to the critical work, the exclusive work
of the current node is equal to sum of the work performed by the left
step siblings and the exclusive work performed by the left finish
siblings of the async child~(line 16 in
Figure~\ref{alg.workspan}).
The algorithm adds the spawn site and the exclusive work performed by
the current async node to the node's spawn site list~(lines 20-22 in
Figure~\ref{alg.workspan}).

After the algorithm completes traversing the entire DPST, the root of
the DPST will contain the list of all spawn sites that perform
critical work and their individual contribution to the critical work.
The root node also contains information about the total work performed
by the program, the work that is computed serially by the program, and
the exclusive work performed under the entry function of the
program~(\ie, \texttt{main}).

\vspace{4pt}
\textbf{Aggregating information about a spawn site.}  A single spawn
site may be executed multiple times in a dynamic execution. Hence,
\taskprof aggregates information from multiple invocations of the same
spawn site. \taskprof computes the aggregate information for each
spawn site by performing another bottom-up traversal of the DPST at
the end. When it encounters an async node, \taskprof uses a hash table
indexed by the spawn site associated with the async node and adds the
total work and critical work to the entry. When aggregating this
information, \taskprof has to ensure that it does not double count
work and critical work when recursive calls are executed. In the
presence of recursive calls, a descendant of an async node will have
the same spawn site information as the async node. If we naively add
the descendant's work, it leads to double counting as the work and
critical work of the current async node already considers the
work/critical work of the descendant async node. Hence, when \taskprof
encounters an async node in a bottom-up traversal of the DPST, it
checks whether the descendants of the async node have the same spawn
site information. When a descendant with the same spawn site exists,
it subtracts such a descendant's work and critical work from the entry
in the hash table corresponding to the spawn site. Subsequently,
\taskprof adds the work and the critical work of the current async
node to the hash table.

\vspace{4pt} \textbf{Profile reported to the user.} For each spawn
site in the program, \taskprof presents the work, the critical work,
the asymptotic parallelism, and the percentage of critical work
exclusively done by the spawn site. The asymptotic parallelism of a
spawn site is the ratio of the total work and the critical work
performed by a spawn site. The spawn sites are ordered by the
percentage of critical work exclusively performed by the spawn
site. Figure~\ref{fig:dpst}(b) illustrates the parallelism profile for
the program in Figure~\ref{fig:example} that has the DPST shown in
Figure~\ref{fig:dpst}(a).  If a spawn site has low parallelism and
performs a significant proportion of the critical work, then
optimizing the task spawned by the spawn site may increase the
parallelism in the program. This profile information provides a
succinct description of the parallelism bottlenecks in the program.

\section{Causal Profiling}
\label{sec:causal}
\taskprof reports the set of spawn sites performing critical work to
the user, which highlight the parallelism bottlenecks in the
program. A programmer can consider these spawn sites to be initial
candidates for optimization to reduce serial computation.

\vspace{4pt}
\textbf{Reducing critical work and the impact on parallelism.}
Designing a new optimization or a parallelization strategy that
reduces the critical work typically requires effort and time. A
program may have multiple spawn sites that perform similar amount of
critical work. When a set of spawn sites are parallelized to reduce
critical work, the resultant execution may have new spawn sites whose
critical work is similar to the critical work before the
optimization. In such cases, an optimization to a spawn site
performing critical work may not improve the asymptotic parallelism
in the program.  Hence, programmers would benefit from a causal
profile of program that identifies the improvement in asymptotic
parallelism when certain regions of the code are optimized.

\vspace{4pt}
\textbf{Causal profile with \taskprof.} A causal profile provides
information on improvements in parallelism when certain parts of the
code are parallelized or optimized.  \taskprof proposes a technique to
generate causal profiles for task parallel programs. The programmer
can get an accurate estimate of the improvement in asymptotic
parallelism by reducing the serial work in a region of the program
using \taskprof's causal profile. \taskprof provides such an estimate
even before the programmer has designed a concrete strategy to
parallelize or reduce the serial work in the region of code under
consideration. In summary, a causal profile enables the programmer to
identify parts of the program that really matter in increasing the
asymptotic parallelism. Figure~\ref{fig:dpst}(c) provides the causal
profile for the program in Figure~\ref{fig:example} where the regions
under consideration are demarcated by \texttt{\_\_CAUSAL\_BEGIN\_\_}
and \texttt{\_\_CAUSAL\_END\_\_}.  Next, we describe how \taskprof
generates a causal profile leveraging the accurate performance model
of a task parallel execution created with the fine-grained attribution
of work and the DPST.

\vspace{4pt}
\textbf{Static code annotations.}
To generate causal profiles, the programmer annotates a static region
of code that is considered for parallelization and the expected
improvement to the critical work from parallelization. The programmer
can provide multiple regions as candidates for optimization. \taskprof
generates a causal profile that estimates the improvement in
parallelism when all annotated regions are optimized. In addition,
\taskprof also generates a causal profile for optimizing each region
in isolation.
Figure~\ref{fig:example} illustrates the regions of code annotated for
causal profiling with \texttt{\_\_CAUSAL\_BEGIN\_\_} and
\texttt{\_\_CAUSAL\_END\_\_} annotations. If the programmer does not
specify the amount of expected improvement for the considered region,
\taskprof assumes a default value. If the annotations are nested, the
outermost region of code is considered for estimating the benefits.

\vspace{4pt} \textbf{Profile execution and attribution of work.}
\taskprof uses these annotations, profiles the program, constructs the
DPST to attribute work to various regions, and provides the estimated
improvement in asymptotic parallelism from optimizing the annotated
regions.  During profile execution, \taskprof measures the work
performed in the annotated part of the step node and also in parts of
the step node that have not been annotated. Hence, each step node can
have multiple work measurements corresponding to static regions with
and without annotation. \taskprof accomplishes it by reading the
performance counter value at the beginning and the end of the each
dynamic region. \taskprof maintains a list of work values for each
step node and writes it to the profile data file.

\vspace{4pt}
\textbf{Algorithm to generate causal profiles.}
The algorithm to compute the causal profile is similar to the work and
span algorithm in Figure~\ref{alg.workspan}. It takes the DPST as
input and a list of anticipated improvements for the annotated
regions. The algorithm outputs a causal profile that computes the
improvement in asymptotic parallelism of the whole program for the
specified improvements of the annotated regions.  The causal profile
algorithm performs a bottom up traversal of the DPST similar to the
work and span algorithm in Figure~\ref{alg.workspan}. However, the
causal profiling algorithm does not track spawn sites and computes the
whole program's work and critical work.  The key difference with the
causal profiling algorithm is the manner in which it handles the work
done by the step nodes, which have regions corresponding to user
annotations. Specifically, \taskprof maintains a list of annotated and
non-annotated regions executed with each step node and the amount of
work performed in each region. To estimate the effect of
optimizing/parallelizing the annotated region, we reduce the critical
work contribution of the annotated region by the user-specified
optimization factor while keeping the total work performed by the
regions unchanged.
The output of the causal profiling algorithm is a list that provides
the asymptotic parallelism for each anticipated improvement factor for
the regions under consideration.

\vspace{4pt} \textbf{Illustration. } After analyzing the parallelism
profile in Figure~\ref{fig:dpst}(b) for the program in
Figure~\ref{fig:example}, the programmer has identified two regions of
code~(lines 4-6 and lines 20-22 in Figure~\ref{fig:example}) for
optimization. The regions are annotated with
\texttt{\_\_CAUSAL\_BEGIN\_\_} and \texttt{\_\_CAUSAL\_END\_\_}
annotations to demarcate the beginning and the end.  During execution,
the region at lines 20-22 is executed once and is represented by step
node $S0$ in Figure~\ref{fig:dpst}(a). In contrast, the
region at lines 4-6 is executed twice and is represented by step nodes
$S5$ and $S6$ in Figure~\ref{fig:dpst}(a). In this
example, the entire step node corresponds to the annotated region. In
general, a step node may have multiple annotated and non-annotated
regions. To generate a causal profile, the
critical work performed by nodes $S0$, $S5$, and $S6$ are decreased by
2$\times$, 4$\times$, and 8$\times$ and its impact on whole program
parallelism is computed. Figure~\ref{fig:dpst}(c) provides the causal
profile with the annotated regions, which reports that the asymptotic
parallelism in the program increases when those two regions are
optimized.

\section{Experimental Evaluation}
\label{sec:eval}
This section describes our prototype, our experimental setup, and an
experimental evaluation to answer the following questions: (1)~Is
\taskprof effective in identifying parallelism bottlenecks?  (2)~Is
\taskprof's parallel profile execution faster than serial
profilers?  (3)~Is \taskprof effective in minimizing perturbation in
the profile execution?  (4)~Is \taskprof usable by programmers?

\vspace{4pt}
\textbf{Prototype.}  We have built a \taskprof prototype to profile
task parallel programs using the Intel Threading Building Blocks(TBB)
library~\cite{Reinders:2007}. The prototype provides a TBB library
that has been modified to construct the DPST, measure work done in
step nodes using hardware performance counters, and track file name
and line information at each spawn site. The prototype also handles
algorithms for geometric decomposition such as \texttt{parallel\_for}
and \texttt{parallel\_reduce}. The prototype also includes a Clang
compiler pass that automatically adds line number and file name
information to the TBB library calls, which enables the programmer to
use the modified library without making any source code
changes. Hence, the modified TBB library can be linked to any TBB
program. Our prototype adds approximately 2000 lines of code to the
Intel TBB library to perform various profiling operations.  The
\taskprof prototype is open source~\cite{taskprof-git}.

\begin{table}
\begin{center}
\begin{adjustbox}{max width=\linewidth}
\begin{small}
\begin{tabular} {| p{1.5cm} | p{2.9cm} | p{0.8cm} | p{0.8cm} | p{0.9cm} | p{1.3cm} |}
  \hline
\centering Application & \centering Description & \centering Speedup & \centering Parallel-ism &  \centering \# of regions &  \centering Causal parallelism \tabularnewline \hline
blackscholes & Stock option pricing & \centering 1.09 & \centering 1.14 & \centering 2 & \centering 59.24 \tabularnewline 
bodytrack & Tracking of a human body & \centering 5.96 & \centering 22.19 & \centering 1 & \centering 40.32 \tabularnewline 
fluidanimate & Simulate fluid dynamics & \centering 9.39 & \centering 66.09 & \centering 1 & \centering 90.2 \tabularnewline 
streamcluster & Clustering algorithm & \centering 7.3 & \centering 55.13 & \centering 2 & \centering 198.93 \tabularnewline 
swaptions & Price a portfolio & \centering 8.59 & \centering 73.45 & \centering 1 & \centering 98.73 \tabularnewline 
convexHull & Convex hull & \centering 1.3 & \centering 1.28 & \centering 4 & \centering 112.17 \tabularnewline 
delRefine & Delaunay Refinement & \centering 2.93 & \centering 5.5 & \centering 7 & \centering 61.28 \tabularnewline 
delTriang & Delaunay triangulation & \centering 1.23 & \centering 1.47 & \centering 5 & \centering 78.85 \tabularnewline 
karatsuba & Karatsuba multiplication & \centering 5.22 & \centering 23.69 & \centering 1 & \centering 36.9 \tabularnewline 
kmeans & K-means clustering & \centering 2.54 & \centering 4.18 & \centering 6 & \centering 69.6 \tabularnewline 
nearestNeigh & K-nearest neighbors & \centering 4.54 & \centering 12.41 & \centering 2 & \centering 30.55 \tabularnewline 
rayCast & Triangle intersection & \centering 6.62 & \centering 48.49 & \centering 2 & \centering 68.52 \tabularnewline 
sort & Parallel quicksort & \centering 3.91 & \centering 6.33 & \centering 2 & \centering 45.04 \tabularnewline 
compSort & Generic sort & \centering 4.99 & \centering 38.97 & \centering 4 & \centering 86.23 \tabularnewline 
intSort & Sort key-value pairs & \centering 4.71 & \centering 48.68 & \centering 2 & \centering 75.02 \tabularnewline 
removeDup & Remove duplicate value & \centering 6.04 & \centering 54.91 & \centering 3 & \centering 98.24 \tabularnewline 
dictionary & Batch dictionary opers & \centering 5.13 & \centering 38.1 & \centering 4 & \centering 73.12 \tabularnewline 
suffixArray & Sequence of suffixes & \centering 3.75 & \centering 5.5 & \centering 1 & \centering 28.53 \tabularnewline 
bFirstSearch & Breadth first search & \centering 6.6 & \centering 22.45 & \centering 5 & \centering 60.55 \tabularnewline 
maxIndSet & Maximal Independent Set & \centering 5.48 & \centering 16.46 & \centering 5 & \centering 52.23 \tabularnewline 
maxMatching & Maximal matching & \centering 6.73 & \centering 46.04 & \centering 0 & \centering 46.04 \tabularnewline 
minSpanForest & Minimum spanning forest & \centering 3.47 & \centering 7.99 & \centering 2 & \centering 49.78 \tabularnewline 
spanForest & Spanning tree or forest & \centering 7.46 & \centering 44.04 & \centering 1 & \centering 58.91 \tabularnewline \hline
\end{tabular}
\end{small}
\end{adjustbox}
\end{center}
\caption{Applications used to evaluate \taskprof. We provide a
  short description of the application, the speedup obtained on a
  16-core machine when compared to serial execution time, the
  asymptotic parallelism reported by \taskprof, the number of
  annotated regions in the program that provides maximum parallelism
  with causal profiling, and the asymptotic parallelism when the
  critical work in the annotated regions is optimized by 100$\times$,
  which we list as causal parallelism.}
\label{tab:benchmarks}
\end{table}

\vspace{4pt} \textbf{Applications used for evaluation.}  We evaluated
\taskprof using a collection of twenty three TBB applications, which
include fifteen applications from the problem based benchmark suite
(PBBS)~\cite{Shun:2012}, all five TBB applications from the Parsec
suite~\cite{bienia08characterization}, and three TBB applications from
the structured parallel programming book~\cite{McCool:2012}.
The PBBS applications are designed to compare different parallel
programming methodologies in terms of performance and code.
We conducted all experiments on a 2.1GHz 16-core Intel x86-64 Xeon
server with 64 GB of memory running 64-bit Ubuntu 14.04.3.
We measured wall clock execution time by running each application five
times and use the mean of the five executions to report performance.
We use the \texttt{perf events} module in Linux to programmatically
access hardware performance counters.

\vspace{4pt}
\textbf{RQ1: Is \taskprof effective in identifying parallelism
  bottlenecks?}
We used \taskprof to identify parallelism bottlenecks in all the 23
applications. Table~\ref{tab:benchmarks} provides details on
applications used, their speedup on a 16-core machine compared to
serial execution, the asymptotic parallelism reported by \taskprof,
the number of regions that we identified using \taskprof to increase
asymptotic parallelism, and the resultant asymptotic parallelism from
causal profiling when the critical work in the identified regions is
decreased by $100\times$.  Typically, asymptotic parallellism of a
program should be at least 10$\times$ or more than the anticipated
speedup on a machine to account for scheduling
overheads~\cite{Frigo:1998, Reinders:2007, Schardl:2015}.

\taskprof's profile shows that some applications in
Table~\ref{tab:benchmarks} have reasonable asymptotic parallelism,
which accounts for a reasonable speedup on a 16-core machine. For
example, \texttt{fluidanimate} application has an asymptotic
parallelism of 66.09 which is the maximum possible speedup when the
program is executed on a large number of machines. The
\texttt{fluidanimate} application exhibits a speedup of 9.39$\times$
compared to a serial execution when the program was executed on a
16-core machine.

Table~\ref{tab:benchmarks} also shows that we were able to identify a
small number of code regions which when optimized provide a
significant increase in asymptotic parallelism. \taskprof's profile
information on spawn sites performing critical work and the causal
profiling strategy was instrumental in identifying the specific
regions of code as candidates for increasing asymptotic
parallelism. The application \texttt{maxMatching} already had a large
amount of asymptotic parallelism and we could not find any region that
increases parallelism.  We were not aware of these parallelism
bottlenecks even though these applications have been widely used,
which emphasizes the need for \taskprof.

\begin{figure}
  \includegraphics[width=\linewidth]{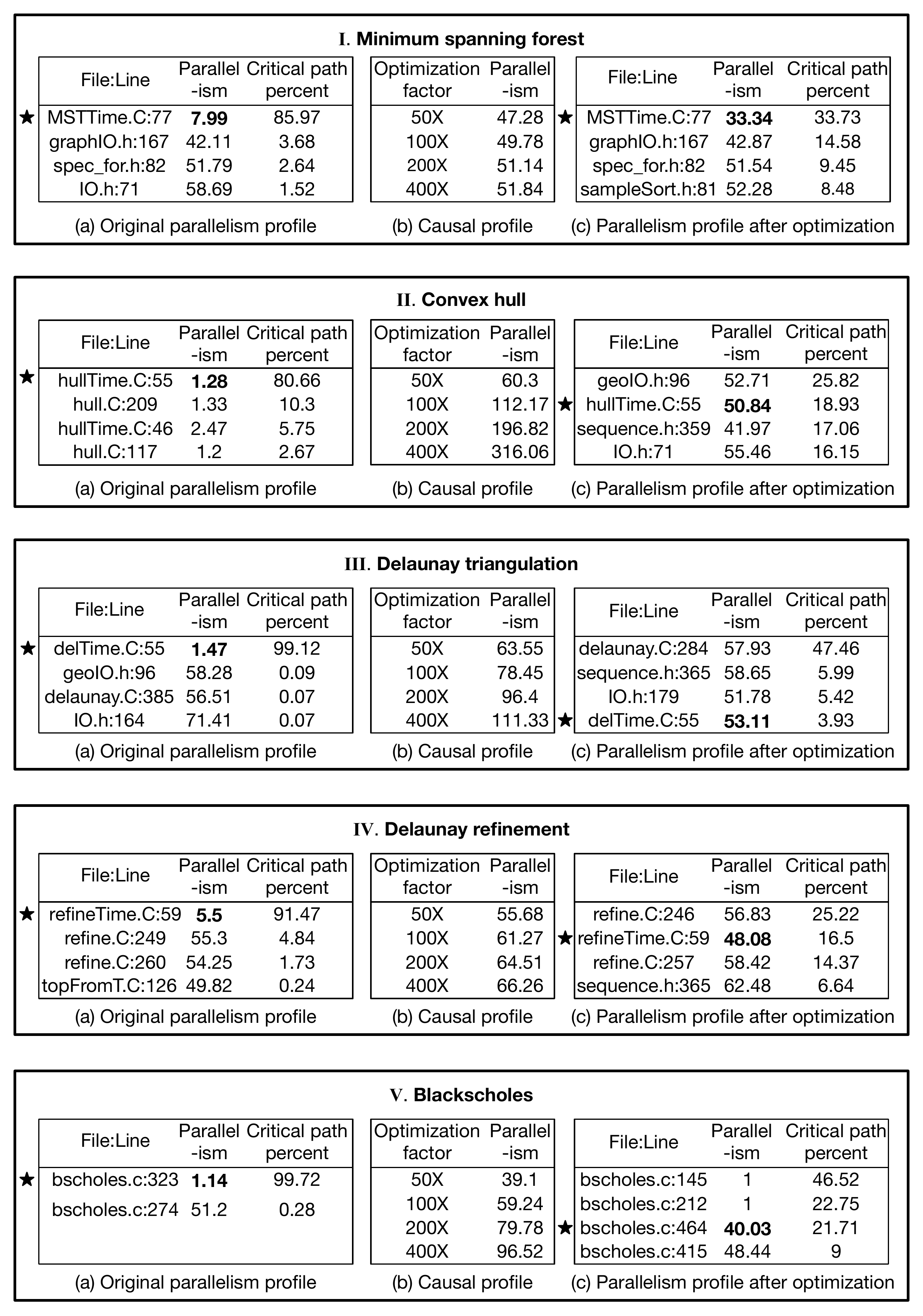}
  \caption{Figure reports the original parallelism profile, the
    causal profile for the annotated regions, and final parallelism
    profile generated by \taskprof after annotated regions were
    parallelized for each of the five applications. We list the top
    four spawn sites from \taskprof's parallelism profile.  Line with
    a ``$\star$'' in the profile corresponds to the main function and
    reports the parallelism for the entire program. The asymptotic
    parallelism for the entire program is marked bold in the
    parallelism profile.}
  \label{fig:all-profs}
\end{figure}

In summary, \taskprof enabled us to identify a set of code regions
that can increase asymptotic parallelism significantly in almost all
our applications. Once we identified code regions that can increase
asymptotic parallelism, we designed concrete parallelization
strategies to reduce the critical work for five applications, which
increased the asymptotic parallelism and the speedup of
the program. We describe them below.

\vspace{4pt}
\textbf{Improving the speedup of the MinSpanningForest application.}
This PBBS application computes the minimum spanning forest of the
input undirected graph. The program has a speedup of 3.47$\times$ over
serial execution on a 16-core machine. The parallelism profile
generated by \taskprof is shown in Figure~\ref{fig:all-profs}(I)(a),
which reports that the parallelism in the program~(main function at
\texttt{MSTTime.C:77}) is 7.99. The main function performs 85\% of the
serial work in the program.
We identified two regions of code using annotations for causal
profiling in the main function. Figure~\ref{fig:all-profs}(I)(b)
presents the causal profile generated by \taskprof, which shows the
increase in asymptotic parallelism in the program on potentially
optimizing these two regions.
On further investigation of the code regions, we realized that
annotated regions were performing a serial sort. We replaced them with
a parallel sort function, which increased the asymptotic parallelism
to 33.34 from 7.99. Figure~\ref{fig:all-profs}(I)(c) reports the
profile after our parallel sort optimization. The speedup of the
program increased from 3.49$\times$ to 6.37$\times$.

\vspace{4pt}
\textbf{Improving the speedup of the Convex Hull application.}
This PBBS application computes the convex hull of a set of points
using a divide and conquer approach~\cite{Barber:1996}.
\taskprof's profile shown in Figure~\ref{fig:all-profs}(II)(a) reveals
that the program has an asymptotic parallelism of 1.28 for the whole
program. As expected, it did not exhibit any
speedup. Figure~\ref{fig:all-profs}(II)(a) shows that 80\% of the
critical work is performed by the spawn site at
\texttt{hullTime.C:55}. We annotated two regions corresponding to that
spawn site, which performed sequential read and write operations of
the input and output files respectively. \taskprof's causal profile
showed that it would increase the parallelism to 6.85. Subsequently,
we annotated two additional regions of code corresponding to the spawn
site performing the next highest critical work (\texttt{hull.C:209})
in Figure~\ref{fig:all-profs}(II)(a). The causal profile shown in
Figure~\ref{fig:all-profs}(II)(b) shows that asymptotic parallelism
increases significantly when all the four regions are optimized.  We
parallelized a loop at spawn site \texttt{hull.C:209} using
\texttt{parallel\_for} and parallelized I/O at spawn site
\texttt{hullTime.C:55}. These optimizations increased the parallelism
to 50.84 (see Figure~\ref{fig:all-profs}(II)(c)) and the speedup of
the whole program increased from 1.3$\times$ to 8.14$\times$.

\vspace{4pt} \textbf{Improving the speedup of Delaunay Triangulation.}
This PBBS application produces a triangulation given a set of points
such that no point lies in the circumcircle of the triangle.
The program has an asymptotic parallelism of 1.47~(see
Figure~\ref{fig:all-profs}(III)(a)) for the entire program and
exhibits little speedup.
The spawn site at \texttt{delTime.C:55} performs 99\% of the critical
work. When we looked at the source code, we found that the program is
structured as a collection of \texttt{parallel\_for} constructs
interspersed by serial code. We annotated five regions of code between
the invocations of \texttt{parallel\_for}. The causal profile in
Figure~\ref{fig:all-profs}(III)(b)) shows that the asymptotic
parallelism increases significantly by optimizing the annotated
regions. We parallelized the annotated regions, which had serial for
loops, using \texttt{parallel\_for} while ensuring they operate on
independent data. The profile for the resultant program is shown in
Figure~\ref{fig:all-profs}(III)(c). The parallelism increased to 53.11
and the speedup increased from 1.23$\times$ to 5.82$\times$.

\vspace{4pt} \textbf{Improving the speedup of Delaunay Refinement.}
This PBBS application takes a set of triangles that form a delaunay
triangulation and produces a new triangulation such that no triangle
has an angle less than a threshold value. \taskprof's profile for this
program reports an asymptotic parallelism of 5.5~(see
Figure~\ref{fig:all-profs}(IV)(a)) and it had a speedup of
2.93$\times$. Similar to delaunay triangulation, this program also had
a set of serial code fragments in-between \texttt{parallel\_for}
calls. We identified seven regions of such serial code and annotated
them. \taskprof's causal profile shown in
Figure~\ref{fig:all-profs}(IV)(b) indicates that optimizing all these
seven regions can increase asymptotic parallelism. We parallelized the
serial for loops in these seven regions using \texttt{parallel\_for},
which increased the asymptotic parallelism to 48.08~(see
Figure~\ref{fig:all-profs}(IV)(c)) and the speedup increased from
2.93$\times$ to 6.42$\times$.




%
\vspace{4pt}
\textbf{Improving the speedup of Blackscholes.}
This application from the PARSEC suite~\cite{bienia08characterization}
computes the price of a portfolio of options using partial
differential equations.
It has low asymptotic parallelism for the entire program~(see
Figure~\ref{fig:all-profs}(V)(a)). This program has a single
\texttt{parallel\_for} that has reasonable parallelism of
51.2. However, the spawn site at \texttt{bscholes.c:323} is performing
99\% of the program critical work. Our examination of the code
revealed that it was reading and writing serially. We split the input
and output into multiple files and parallelized the input/output
operations which increased the parallelism to 40.03 and the speedup
increased from 1.09$\times$ to 7.7$\times$.

In summary, \taskprof enabled us to quantify asymptotic parallelism in
the program and its causal profiling strategy enabled us to identify
specific regions of code that can increase parallelism.
 
\begin{figure}
  \centerline{\includegraphics[width=\linewidth]{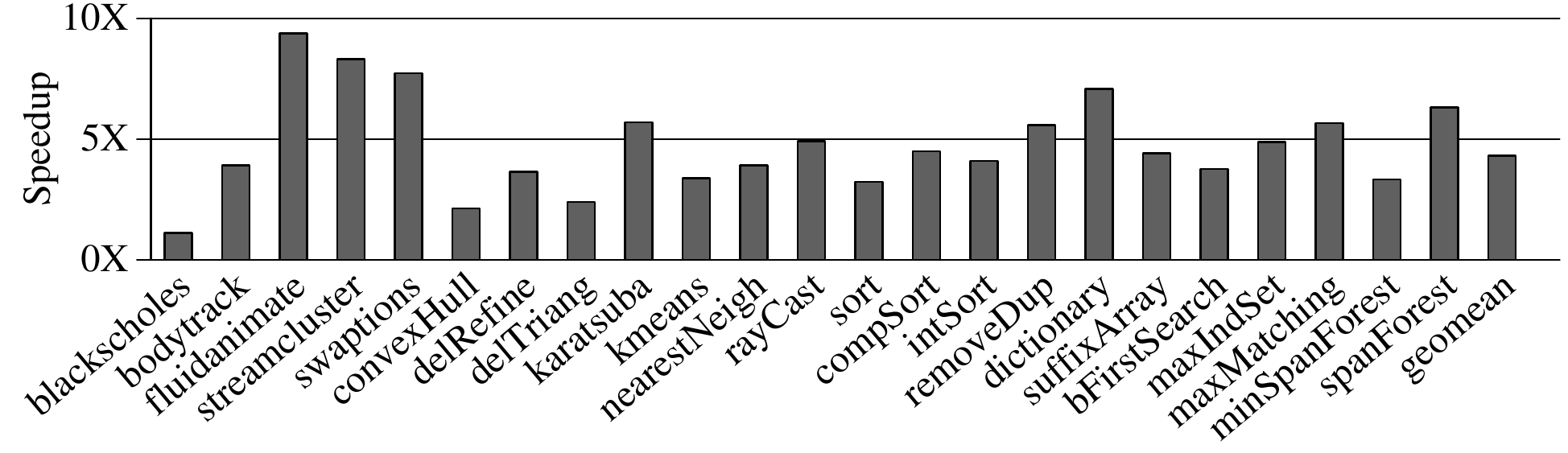}}
  \caption{Speedup of \taskprof's parallel profile execution when
    compared to serial profile execution.}
  \label{fig:speedup}
\end{figure}

\vspace{4pt}
\textbf{RQ2: Is \taskprof's parallel profile execution faster than
  serial profile execution?}
\taskprof's profile execution executes in parallel compared to prior
profilers~\cite{Schardl:2015}, which execute serially. To quantify the
benefits of parallel profile execution, we designed a serial version
of \taskprof by pinning the execution of the program to a single core. 
This is an approximation of serial profiling as TBB programs do not
have serial semantics. Figure~\ref{fig:speedup} reports the speedup of
a parallel \taskprof profile execution compared to a serial profile
execution. On average, \taskprof's parallel profile execution is
4.32$\times$ faster than serial profile execution. The speedup from a
parallel profile execution is proportional to the amount of
parallelism in the application.


\vspace{4pt} \textbf{RQ3: Is \taskprof effective in minimizing
  perturbation in the profile execution?}
\taskprof uses hardware performance counters to perform fine-grain
attribution of work and to minimize perturbation. The average
performance overhead of \taskprof's profile execution compared to the
parallel execution of the program without any profiling
instrumentation is 56\%. A major fraction of this performance overhead
is attributed to system calls to read hardware performance
counters. \taskprof's profile execution is an order of magnitude
faster than instrumenting each dynamic instruction through compiler
instrumentation, which exhibited overheads of 20$\times$-100$\times$
for the applications in Table~\ref{tab:benchmarks}. Hence, \taskprof
minimizes perturbation even with fine-grained attribution of work.

\vspace{4pt}
\textbf{RQ4: Is \taskprof usable by programmers?}
We conducted a user study to evaluate the usability of \taskprof. The
user study had thirteen participants: twelve graduate students and one
senior undergraduate student. Among them, two students had 4+ years of
experience in parallel programming, five students had some prior
experience, four students had passing knowledge, and two students had
no prior experience with parallel programming. The total duration of
the user study was four hours. To ensure that every student had some
knowledge in parallel programming, we provided a 2-hour tutorial on
task parallelism, and on writing and debugging task parallel
programs using Intel TBB. We gave multiple examples to demonstrate
parallelism bottlenecks.

After the tutorial, the participants were given a total of four
applications and were asked to identify parallelism bottlenecks
without using \taskprof in a one hour time period. Three applications
--- minSpanForest, convexHull, and blackscholes --- from
Table~\ref{tab:benchmarks} and a treesum application similar to the
example in Figure~\ref{fig:example}. We chose these applications as it
had varying levels of difficulty in diagnosing parallelism
bottlenecks.  We asked the participants to identify the static region
of code causing the bottleneck and record the time they spent to
analyze each program. They were not required to design any
optimization.  Some participants used \texttt{gprof} and others used
fine-grained wall clock based timing for assistance.  At the end of
the time period, twelve of them did not correctly identify
parallelism bottlenecks in any of the four applications. One
participant, who had 4+ years of experience in parallel programming,
identified the bottleneck in one (minSpanForest) out of the four
applications. Some participants were misled by the \texttt{gprof}
profile.

Subsequently after the first part, we gave a brief tutorial
of \taskprof on a simple example program. The participants were then
asked to identify bottlenecks in the four applications using \taskprof
within an hour. Using \taskprof, seven participants found the
parallelism bottleneck in all the four applications, one participant
found the bottleneck in three of them, four participant found the
bottleneck in two of them, and one participant did not find the bottleneck in any
application. Among the participants who identified at least one
bottleneck for any application, it took them 12 minutes on average per
application to identify the bottleneck using \taskprof. The
participants indicated that once they became familiar with the tool by
identifying a bottleneck in one application, subsequent tasks were
repetitive.  In summary, our user study suggests that \taskprof can
enable both expert and relatively inexperienced programmers identify
parallelism bottlenecks quickly.

\section{Related Work}
There is a large body of work to identify parallelism
bottlenecks. These include techniques to address load
imbalances~\cite{Tallent:2010:SIL,DeRose:2007,Oh:2011,Kambadur2014},
scalability bottlenecks~\cite{Tallent:2009,Schardl:2015,Liu:2015},
visualizing bottlenecks~\cite{DuBois:2013, DuBois:2013:CSI,
  Eyerman:2012, knupfer2008vampir}, synchronization
bottlenecks~\cite{Chen:2012,David:2014,Yu:2016}, and data locality
bottlenecks~\cite{Liu:2011,Liu:2013:PDL,Acar:2000}. Data locality and
synchronization bottlenecks increase serial work. Hence, \taskprof
will report asymptotic parallelism in their presence.  In contrast to
prior proposals, \taskprof also estimates the
improvement in parallelism with causal profiling.  Next, we focus on
the closest related work.

\vspace{4pt}
\textbf{Profiling tools for task parallel programs.}
Profiling tools such as HPCToolkit~\cite{Adhianto:2010}, and Intel
VTune Amplifier~\cite{vtuneAmplifier} can analyze a program's
performance on various parameters using hardware performance counters.
HPCToolKit also has metrics to quantify idleness and the scheduling
overhead~\cite{Tallent:2009} in Cilk programs that is specific to a
machine. They do not compute the asymptotic parallelism in the
program. They also do not identify code that matters with respect to
asymptotic parallelism.
%
%
CilkView~\cite{He:2010} computes the whole program asymptotic
parallelism. CilkProf~\cite{Schardl:2015} computes asymptotic
parallelism per spawn site using an online algorithm. However, these
profilers execute the program serially, which is only possible with
Cilk programs with C-elision~\cite{Frigo:1998}. Many task parallelism
frameworks including Intel TBB do not have serial semantics, which
limits their use.
Further, executing the profiler serially can cause high
overheads. Unlike \taskprof, they also cannot estimate the benefits of
optimizing specific regions of code.

\vspace{4pt} \textbf{Performance estimation tools.}  An early
profiling technique proposed Slack~\cite{Hollingsworth:1994}, which is
a metric that estimates the improvement in execution time through
critical path optimizations for a specific machine model.
Kremlin~\cite{Garcia:2011}
identifies regions of code that can be parallelized in serial programs
by tracking loops and identifying dependencies between iterations.
Kismet~\cite{Jeon:2011} builds on Kremlin to estimate speedups for the
specific machine on which the serial program is executed.  These
techniques are tied to a specific machine and cannot estimate
asymptotic parallelism improvements.

Our work is inspired by \textsc{Coz}~\cite{Curtsinger:2015}, a causal
profiler for multithreaded programs that automatically identifies
optimization opportunities and quantifies their impact on a metric of
interest, such as latency or throughput. It runs periodic experiments
at runtime that virtually speed up a single randomly selected program
fragment. Virtual speedups produce the same effect as real speedups by
uniformly slowing down code executing concurrently with the fragment,
causing the fragment to run relatively faster.
In a task parallel context, it is not possible to slow down all active
tasks. Further, slowing down threads does not measure the impact of
the region as work stealing dynamically balances the load.  Further,
\textsc{Coz}'s virtual speedups are specific to a particular
machine. \taskprof, though similar in spirit, addresses the above
challenges and proposes a causal profiler that leverages the dynamic
execution structure and estimates improvements in asymptotic
parallelism. Hence, \taskprof's profile is not specific to a single
machine and enables the development of performance portable code.

\section{Conclusion}

\taskprof identifies parallelism bottlenecks by performing a
low-overhead, yet fine-grained attribution of work to various parts of
the program using the dynamic execution structure of a task parallel
execution. \taskprof reports asymptotic parallelism and serial work
performed at each spawn site. \taskprof's causal profile estimates the
improvements in parallelism when regions of code annotated by the
programmer are optimized.  We have identified bottlenecks and improved
the speedup in numerous Intel TBB applications. Our user study shows
that developers can quickly identify parallelism bottlenecks using
\taskprof.

\begin{acks}
We thank the anonymous reviewers for their feedback. This paper is
based on work supported in part by NSF CAREER Award CCF--1453086, a
sub-contract of NSF Award CNS--1116682, and a NSF Award CNS--1441724.
\end{acks}

\bibliographystyle{ACM-Reference-Format}
\balance
\bibliography{concurrency}
\appendix
\end{document}